# Observation of intervalley quantum interference in epitaxial monolayer WSe$_2$

H. J. Liu[1], J. L. Chen[1], H. Y. Yu[1,2], F. Yang[3,4], L. Jiao[1], G. -B. Liu[5], W. K. Ho[1], C. L. Gao[3,4], J. F. Jia[3,4], W. Yao[1,2], M. H. Xie[1*]

[1]*Physics Department, The University of Hong Kong, Pokfulam Road, Hong Kong*

[2]*Center of Theoretical and Computational Physics, The University of Hong Kong, Pokfulam Road, Hong Kong*

[3]*Key Laboratory of Artificial Structures and Quantum Control (Ministry of Education), Department of Physics and Astronomy, Shanghai Jiaotong University, 800 Dongchuan Road, Shanghai 200240, China*

[4]*Collaborative Innovation Center of Advanced Microstructures, Department of Physics and Astronomy, Shanghai Jiao Tong University, Shanghai 200240, P. R. China*

[5]*School of Physics, Beijing Institute of Technology, Beijing 100081, China*

**Monolayer (ML) transition metal dichalcogenides (TMDs) have been attracting great research attentions lately for their extraordinary properties, in particular the exotic spin-valley coupled electronic structures that promise future spintronic and valleytronic applications[1-3]. The energy bands of ML TMDs have well separated valleys that constitute effectively an extra internal degree of freedom for low energy carriers[3-12]. The large spin-orbit coupling in the TMDs makes the spin index locked to the valley index, which has some interesting consequences such as the magnetoelectric effects in 2H bilayers[13]. A direct experimental characterization of the spin-valley coupled electronic structure can be of great interests for both fundamental physics and device applications. In this work, we report the first experimental observation of the quasi-particle interference (QPI) patterns in ML WSe$_2$ using low-temperature (LT) scanning tunneling microscopy/spectroscopy (STM/S). We observe intervalley quantum**

**interference involving the Q-valleys in the conduction band due to spin-conserved scattering processes, while spin-flip intervalley scattering is absent. This experiment establishes unequivocally the presence of spin-valley coupling and affirms the large spin-splitting at the Q valleys. Importantly, the inefficient spin-flip intervalley scattering implies long valley and spin lifetime in ML WSe$_2$, which represents a key figure of merit for valley-spintronic applications.**

Ultrathin WSe$_2$ films are grown on highly ordered pyrolytic graphite (HOPG) by molecular-beam epitaxy (MBE). The electronic structure and QPI in ML WSe$_2$ are probed by STM/S at 77 K. LT-STS is a known powerful method for probing electronic structures of thin films and its studies have already been implemented to extract the quasiparticle band gaps and band edges in ML TMDs[14-18]. Quantum interference has been also studied by STM/S for metals[19], topological insulators[20,21] and graphene[22]. A defect elastically scatters electrons from wave vector $\vec{k}_i$ to $\vec{k}_f$ on the constant energy contour (CEC), and quantum interference of the incoming and scattered waves form standing waves of wavevector $\vec{q} = \vec{k}_f - \vec{k}_i$, which are detected by LT-STS mapping. For multi-valley bands such as that in graphene and ML-TMDs, the scattering by a point defect can give rise to rich QPI patterns arising from quantum interference within a valley as well as between well-separated inequivalent valleys. Fourier transformation (FT) of the STS maps (FT-STS) can then directly reveal the valley locations and the band dispersion relations[23,24], for example.

**Fig. 1a** shows a topographic image of an as-grown WSe$_2$ sample. It reveals the characteristic terrace-and-step morphology of an atomically flat film. The sample has the nominal coverage of 1.2 MLs, so in addition to the ML film over large surface areas, there are also bilayer and even trilayer high islands. Comparing to the MBE-grown MoSe$_2$ films[18], the most striking feature in Fig. 1a is the absence of inversion domain boundary network, which is commonly seen in epitaxial MoSe$_2$. This has given us an opportunity to study the quantum interference effect in ultrathin

WSe$_2$ by STM/S, thereby allows probing the electronic structure as well as the spin-splitting in ML-TMDs. The inset in Fig. 1a is a close-up atomic resolution image of the ML region of the sample. Fig. 1b shows the differential conductance spectrum measured at a fix point on ML WSe$_2$ by STS at 77 K. It reveals an energy gap of 2.59 ± 0.07 eV, in agreement with the previous report[16]. The Fermi level is found slightly above the mid-gap energy, suggesting the sample is slightly electron-doped, likely by some native defects such as Se vacancies.

To search for the QPI in ML WSe$_2$, we firstly locate a surface area that contains point defects. An example is given in Fig. 1c, which shows two point defects in the field of view. In this region of the surface, quantum interference stands high chance to be observed by LT-STM/S. We began our search of the QPIs by taking the STS maps at energies close to the valence band maximum (≤ -1.4 eV). Fig. 1d presents one of the STS maps obtained and the inset shows the power spectrum through Fourier transform. For all the STS maps taken in the energy range of -1.4 eV ~ -1.5 eV, including Fig. 1d, no sign of intervalley QPI is seen and the FT-STS only reveal the reciprocal lattice vectors (G) of WSe$_2$. The latter has, nevertheless, allowed one to determine the Brillouin zone (BZ) as marked by the white dashed line in the figure.

The absence of intervalley QPI in the valence band implies that spin-flip scattering between K and $\bar{K}$ valleys is inefficient. In ML WSe$_2$, the ultra-strong spin-orbit coupling in the 5d orbitals of the metal atoms give rise to large spin splitting in both the conduction and valence bands, which are dictated by the mirror symmetry and time-reversal symmetry to be in the opposite out-of-plane directions at a time reversal pair of either K or Q valleys[3]. In Fig. 2a, we show electronic bands of ML WSe$_2$ calculated by the density functional theory (DFT), in which the red solid and dashed blue lines represent the spin-split bands due to spin-orbital coupling. The valence band edge at the K points has a spin splitting of ~ 0.4 eV with opposite signs at K and $\bar{K}$ as required by the time reversal symmetry[17,25]. Consequently quasi-particle interference between K and $\bar{K}$ valleys will be prohibited by the time-reversal symmetry if the scattering defects are nonmagnetic. Moreover, a recent study has shown that STM/S is not very sensitive to the K-valley states as compared to the Γ and Q valleys[17], which can be another cause for the absence of QPI in the STS near the valence band maximum.

Conduction band electron has a completely different story. The spin-splitting is much smaller at the K-point (~ 0.03 eV), and the largest spin-split of ~ 0.2 eV occurs in the Q-valley (*c.f.* Fig. 2a). The energy minimum at Q is close to that of the K valleys in ML WSe$_2$[3,11,17,25]. Besides, Q valley can have a much larger weight in STM/S than K valley due to the larger density of states (DOS) as well as a larger tunneling coefficient[3,17]. Therefore the Q valleys are expected to play a significant role in the STS mapping of the empty-states. As illustrated in Fig. 2b, the six-fold degenerate Q valleys form two groups. $Q_1$, $Q_2$ and $Q_3$ have the same spin and $\bar{Q}_1$, $\bar{Q}_2$ and $\bar{Q}_3$ are their time reversals. QPIs by the spin-conserved intervalley scattering are possible within each group. In addition, as the spin-split is small at K while the energy difference between Q and K is also small, QPIs by the spin-conserved scatterings between K and $\bar{K}$, and between Q and K-valleys may also occur (*c.f.* Fig. 3d).

Figure 3a presents an example of the STS maps obtained over the same surface area as in Fig. 1c but at an energy of *E* = +1.0 eV. Unlike the filled-state STS map of Fig. 1d, the empty-state image reveals clearly the QPI patterns in the vicinity of the two defects. Figs. 3b and 3c show the corresponding FT-STS maps presented in the top and perspective views, respectively, from which one notes distinct intensity spots besides the reciprocal lattice (G), signifying the various scattering channels and the corresponding standing wave vectors $\vec{q}$ as labelled. In order to identify the relevant scattering processes, we show in Fig. 3d a CEC map derived from the DFT calculations. This CEC map corresponds to an energy that is slightly above the conduction-band minimum (the black horizontal line in Fig. 2a), where the spin texture of the Q-valley is clearly resolved. For the K-valley, on the other hand, both spin-up and spin-down bands are present with a small difference on the CEC. Based on this CEC, we may identify a number of possible spin-conserved intervalley scatterings as indicated by the green arrows. A spin-flip inter-Q valley scattering is also marked by a dotted orange line. By comparing with the experimental data, we may assign the strongest scattering with the wave-vector close to the M point of the BZ (i.e., the middle point on the edge of the dashed white hexagon) is of the spin-conserved Q-Q intervalley scattering, i.e., $\vec{q}_4$ in Fig. 3d.

A possible complication, however, is the presence of scattering between Q and K-valleys, which would result in a wavevector ($\vec{q}_3$ in Fig. 3d) that is very similar to $\vec{q}_4$. With the resolution of the experimental data, we cannot discriminate one from the other, and indeed both could occur. Nevertheless, according to the band structure calculations[11,26], the Q-valley is situated slightly off the mid-point of Γ-K and closer to K (see Fig. 2a), so $\vec{q}_4$ would be slightly larger than |ΓM|, which appears consistent with the experimental data. Moreover, as noted earlier, the STM/S measurement should be more sensitive to Q states than the K valleys, therefore the Q-Q QPI has a larger chance to be picked up in the STM/S images compared to the Q-K one. We therefore believe that Q-Q scattering ($\vec{q}_4$) is the dominating contributor to the experimentally observed spot next to M in Figs. 3b and 3c.

With the presence of both spin-up and down CEC at K valleys and the six-fold degenerate Q valleys, Q-K QPI at another two wavevectors, $\vec{q}_2$ and $\vec{q}_5$, and $K - \overline{K}$ QPI at wavevector $\vec{q}_1$ are also possible through the spin-conserved scattering (see Fig. 3d). Though weak, intensity spots are indeed observable at $\vec{q}_2$ and $\vec{q}_5$ as shown in Figs. 3b and 3c. Intensity spot at wavevector $\vec{q}_1$ may not be discernable in Fig. 3b or 3c, which again may be attributed to the sensitivity of STM/S to K-valley electrons. However, we did observe hint of such scatterings at some other energies (see Supplementary).

We have performed calculations of the joint density of states, $JDOS(\vec{q}) = \int I(\vec{k})I(\vec{k}+\vec{q})d^2\vec{k}$, where $I(\vec{k})$ represent the DOS derived from the DFT calculated energy bands, and Fig. 3e and 3f show two such JDOS maps at the energy of Fig. 3d for the spin-conserved and spin-flip scatterings, respectively. For comparison and completeness, Fig. 4 presents another CEC map (4c) and the corresponding JDOS (4d and 4c for the spin-conserved and spin-flip scatterings, respectively) for a higher energy (i.e., the green line in Fig. 2a) and compared with an experimental FT-STS obtained at 1.2 eV (4a and 4b for the top and perspective view, respectively). It is clear that the spin-conserved scattering captures well the experimental FT-STS maps, elucidating the non-magnetic nature of the defect. The agreement between Figs. 3b and 3e also provides an experimental evidence for the presence of an appreciable spin-splitting at Q-valleys in ML WSe$_2$, as otherwise the scattering would resemble that of Fig. 3f, where $\overline{Q}_i - Q_j$ scattering channels would be present (i.e., the black arrow in Fig. 4c). This is clearly not the case in the experiment.

By examining the STS maps obtained at different energies, we derive an energy dependence of the $|\vec{q}|$'s (peak-to-peak distances in FT-STS maps), and Fig. 4f summarizes the data for the wavevector $\vec{q}_4$. There are apparently two branches, corresponding to $|\vec{q}_4| \sim 1.06$ Å$^{-1}$ and 1.14 Å$^{-1}$, respectively. According to first-principle band structures, the exterma of the upper and lower spin-subbands at the Q-valley do show a relative shift in momentum (as indicated by two short vertical arrows in Fig. 2a). Therefore we are tempted to attribute the two branches in Fig. 4f to be scatterings dominantly in the lower and upper spin-subbands, respectively. If this is indeed the case, we may further estimate the magnitude of the spin-split $\Delta_{SO}$ at Q, which is about 0.2 eV. This value compares reasonably well with the first-principle calculation estimations[11]. However, a discrepancy exists, where most first-principle calculations suggest the upper spin-subband at Q has a smaller momentum than the lower one, which is opposite to the finding of Fig. 4f. Besides, the experimental data can have contributions of the other scattering channels (e.g. the Q-K scattering $\vec{q}_3$). Hence the two momentum branches observed in Fig. 4f may reflect a change of dominance of the different scattering processes when the energy is changed. Further studies by higher resolution STM/S may resolve this ambiguity.

Lastly, we draw attention of the seemingly ring feature in the FT-STS map at the center (Fig. 3b and 3c). It may have reflected intra-valley scattering of electrons. Unfortunately because of long wavelength (and thus small $\vec{q}$), the ring feature cannot be separated well from the central bright spot often affected by the random noise of the STS maps. So we have not been able to derive the dispersion relation of the

Q-valley in this experiment.

In conclusion, we have observed for the first time the QPI patterns by intervalley scattering in MBE-grown WSe$_2$ ML. Particularly, they are observed for conduction band electrons. By comparing with the first principle band structures, we establish the dominance of spin-conserved inter Q-valley scattering by point defects in as-grown ML WSe$_2$. The absence of $\overline{Q}_i - Q_j$ scattering affirms large spin-splitting at Q and implies long valley and spin lifetime of the valley electrons, a key figure of merit for valley-spintronic applications. Two branches of the QPI wavevectors are discovered, which may reflect the shift of Q-valley minima between the two spin-subbands. Finally, signature of intra-valley scattering is also noted, which invites future studies for the dispersion relation of the Q-valley conduction band electrons.

## Method

**Film growth**: Ultrathin WSe$_2$ film was grown on freshly cleaved HOPG in a customized Omicron MBE reactor with the base pressure of $10^{-10}$ mbar. Elemental Tungsten (W) of purity of 99.99% and Selenium (Se) of 99.999% in purity were used as the sources, and their fluxes were generated from an e-beam evaporator and a dual-filament Knudsen cell, respectively. A flux ratio of 1:15 between W and Se was adopted. The substrate temperature during film deposition was ~ 400 ºC and the deposition rate was 0.5 MLs/hr. During growth, the film surface was monitored *in situ* by reflection high-energy electron diffraction (RHEED), showing the streaky patterns, signifying two-dimensional (2D) layer-by-layer growth of WSe$_2$ on HOPG. After a preset coverage of the deposit was grown, the source fluxes were cut off by mechanical shutters, in the meantime, the sample temperature was brought to RT by natural cooling upon shutting off the power to the sample manipulator.

**STM measurement:** Room temperature STM measurements were performed in situ using an Omicron VT-STM system and the LT-STM measurements were carried out at 77 K in a Unisoku LT-STM system ex situ. For the latter, the sample surface was protected by an amorphous Se layer deposited at RT until the RHEED pattern became featureless but a diffusive background before being taken out of vacuum. After transferred to the Unisoku LT-STM system, the sample was gently annealed at 300 ºC until the streaky RHEED patterns reappeared and the same flat surface morphology recovered according to STM examinations. For both RT and LT STM measurements, the constant current mode was used; while for STS measurements, the lock-in technique was employed using a modulation voltage of 15 mV and frequency of 985 Hz.

**DFT and JDOS calculations:** The DFT calculations are done by the VASP code[27] using the projector augmented wave[28] as well as the Perdew-Burke-Ernzerhof exchange-correlation functional[29]. Relaxed lattice parameters are used for $WSe_2$ monolayer[30]. Spin-orbit coupling is considered in the calculation. The energy cutoff of the plane-wave basis is set to 400 eV and the energy convergence is $10^{-6}$ eV. Vacuum layer is greater than 15 Å to separate neighboring periodic images. A Γ-centered k-mesh of $10 \times 10 \times 1$ is used to obtain the ground-state density and a 51×51 k-mesh is used to calculate the band energies in the rhombus reciprocal cell.

After getting the conduction band dispersion $E_{\vec{k}}$ of the 2D first Brillouin zone, the $JDOS(\vec{q}) = \int I(\vec{k}) I(\vec{k}+\vec{q}) d^2\vec{k}$ at the Fermi level $E_F$ is calculated using the approximation $I(\vec{k}) = \delta(E_{\vec{k}} - E_F) \approx \frac{1}{\sqrt{\pi}\Delta E} \exp\left(-\left(\frac{E_{\vec{k}}-E_F}{\Delta E}\right)^2\right)$. We set the energy resolution $\Delta E = 30$ meV which well suits our $51 \times 51$ reciprocal cell mesh grid.

**Acknowledgements**

The work described in this paper is supported by a Collaborative Research Fund (HKU9/CRF/13G) sponsored by the Research Grant Council (RGC), Hong Kong Special Administrative Region. M. X. and J. J. acknowledge the support from a grant of the SRFDP and RGC ERG Joint Research Scheme of Hong Kong RGC and the Ministry of Education of China (No. M-HKU709/12). The work in SJTU was also supported by the MOST (2013CB921902, 012CB927401) and the NSFC (11227404, 11374206) of China. G. L. acknowledges the support from the NSFC (11304014) of China.

**Author Contributions**

M. X. and H. L. conceived and designed the experiments; H. L., J. C., F. Y., L. J. and W. H. performed the growth and STM experiments; H. Y., G. L. and W. Y. calculated the energy bands and the JDOS; M. X., W. Y., H. L., H. Y. and J. C. analyzed the data: M. X., C. G. and J. J. contributed the experimental equipment; M. X. and W. Y. co-wrote the paper.

Corresponding author: M.H. Xie ([*mhxie@hku.hk](*mhxie@hku.hk))

**Figures and captions**

**Fig. 1  STM/S of MBE-grown WSe$_2$**  (a) STM topographic image (size: 75 × 75 nm$^2$, sample bias: 2.4 V) of a MBE grown WSe$_2$ sample on HOPG with the nominal coverage of 1.2 MLs, revealing dominantly ML WSe$_2$, bilayer (BL) and trilayer (TL) islands. Inset: an atomic resolution image (size: 7.5 × 7.5 nm$^2$, sample bias: -1.5 V) of ML WSe$_2$. (b) STS spectrum (averaged over 50 scans) taken at a fixed position on a defect-free ML WSe$_2$, revealing an energy gap of 2.59 eV. (c) Topographic STM image and (d) the STS map of ML WSe$_2$ measured simultaneously at -1.5 eV (size: 14 × 14 nm$^2$). The white dashed circles in (c) mark two point defects, which show darker contrast in (d). The inset in (d) is the Fourier transform of the STS map showing the reciprocal lattice (the points labeled by G). The first BZ (the white dashed hexagon) and its high symmetry points (K and M) are also shown for reference.

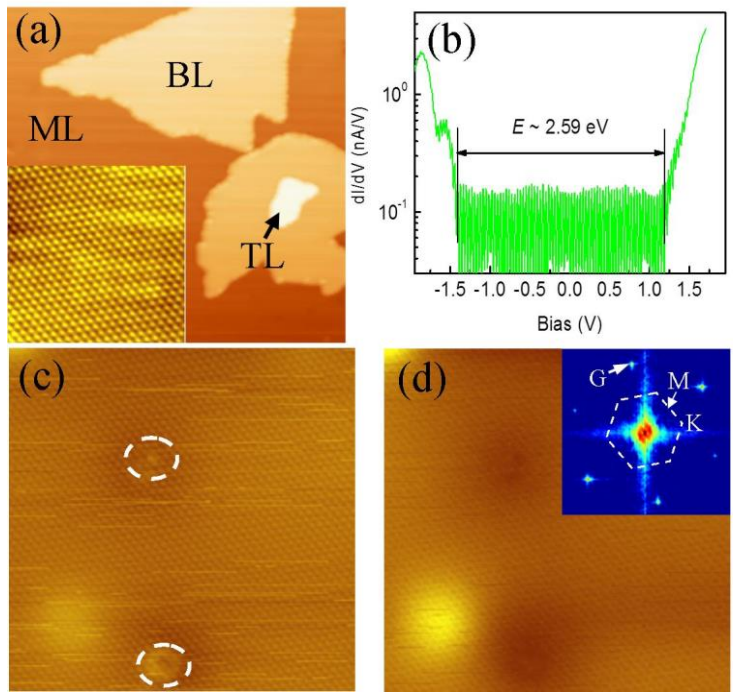

**Fig. 2  K and Q valleys in ML WSe$_2$**    (a) Electronic bands of ML WSe$_2$ calculated by density functional theory. In the bottom conduction and top valence bands, the spin down and up subbands are denoted by red and blue color respectively. The vertical arrows point to the minima of the spin-spilt bands at Q-valley. The two horizontal lines mark the energies at which the constant energy contour maps in Fig. 3d and 4c are obtained. (b) Schematic illustration of the spin-valley coupled conduction band edges, where the blue and red colors denote the spin-up and spin-down bands, respectively.

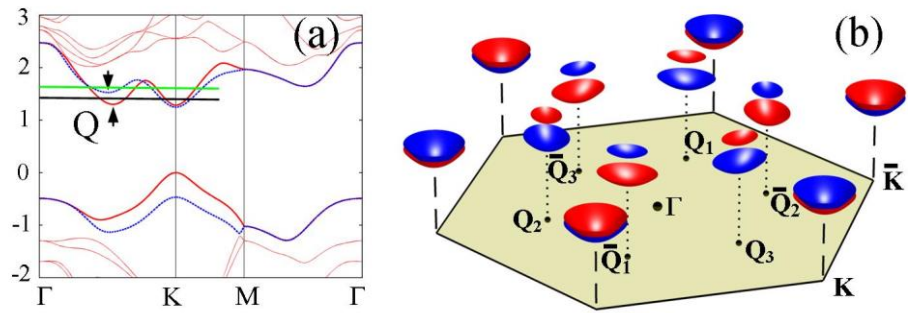

**Fig. 3 Intervalley quantum interference at defects.** (a) STS map of ML $WSe_2$ at +1.0 eV of the same surface as in Fig. 1c, showing QPI patterns in the vicinity of the two point defects. (b, c) The Fourier transform of the STS map in (a), shown in the top and perspective views, respectively. The white dashed hexagon in (b) shows the first BZ for reference. The observed scattering wavevectors ($q_i$'s) are exemplified by arrowed solid white lines in (b) and labelled in (c). The central ring-like feature as highlighted by the yellow dotted circle in (b) may reflect the intravalley scattering. (d) Constant energy contour at the energy marked by the black horizontal line in Fig. 2a. The blue (red) color denotes the up (down) spin states. The solid green arrows indicate five possible spin-conserved scattering channels with the resulting interference wave vectors $q_i$. The dashed orange arrow indicates a spin-flip scattering channel. (e, f) Calculated joint density of states for, respectively, spin-conserved and spin-flip scatterings at the same energy as in (c).

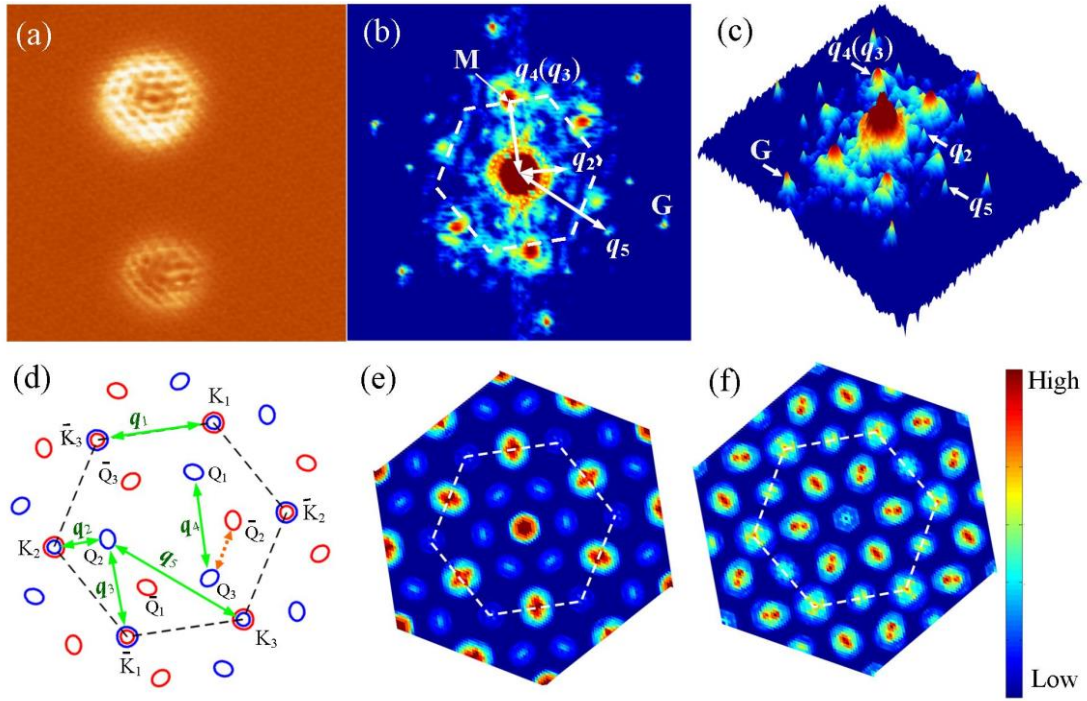

**Fig. 4 Energy dependence of the quantum interference pattern** (a, b) Fourier transform of the STS map of the surface as in Fig. 3 but measured at energy 1.2 eV, shown in the top and perspective views, respectively. (c) Constant energy contour at an energy marked by the horizontal green line in Fig. 2a, where the blue (red) line denotes the up (down) spin states. The solid green arrow indicates a spin-conserved $Q_i - Q_j$ scattering channel. The black arrow indicates spin-conserved $\overline{Q}_i - Q_j$ intervalley scattering channel, as both spin-up and spin-down bands at Q valleys are relevant at this higher energy. (d, e) The joint density of state maps calculated for, respectively, spin-conserved and spin-flip scatterings at same the energy as in (c). (f) Experimentally derived wave vectors ($\vec{q}_4$) from the FT-STS maps at different energies (from 0.7 eV to 1.2 eV), revealing two branches as highlighted by the solid red lines.

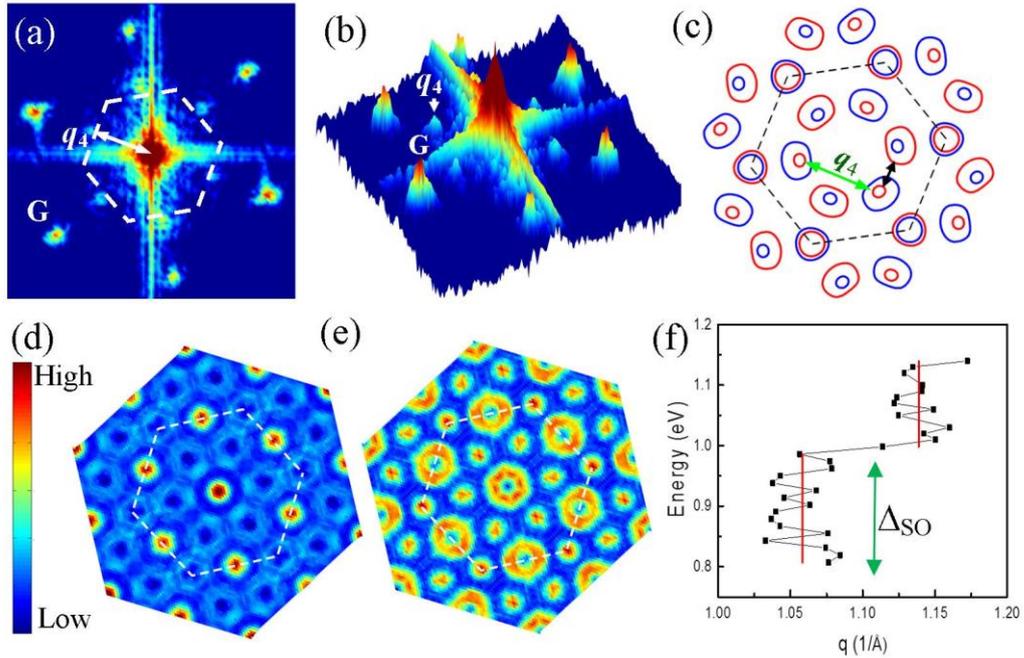